\documentclass[10pt,a4paper]{article}

\newcommand{\xx}{\mathbf{x}}
\newcommand{\xa}{\mathbf{x}^a}
\newcommand{\xb}{\mathbf{x}^b}
\newcommand{\xt}{\mathbf{x}^t}
\newcommand{\yo}{\mathbf{y}^o}

\newcommand{\oA}{\mathbf{A}}
\newcommand{\oB}{\mathbf{B}}
\newcommand{\oC}{\mathbf{C}}

\newcommand{\oH}{\mathbf{H}}
\newcommand{\oI}{\mathbf{I}}
\newcommand{\oK}{\mathbf{K}}

\newcommand{\oM}{\mathbf{M}}

\newcommand{\oP}{\mathbf{P}}
\newcommand{\oR}{\mathbf{R}}
\newcommand{\oS}{\mathbf{S}}
\newcommand{\oT}{\mathbf{S}}
\newcommand{\oQ}{\mathbf{Q}}
\newcommand{\oU}{\mathbf{U}}
\newcommand{\oV}{\mathbf{V}}

\newcommand{\Zero}{\mathbf{0}}

\newcommand{\tP}{\mathbf{\tilde{P}}}
\newcommand{\tR}{\mathbf{\tilde{R}}}
\newcommand{\tS}{\mathbf{\tilde{S}}}
\newcommand{\tQ}{\mathbf{\tilde{Q}}}

\usepackage{graphicx}
\usepackage{dcolumn}
\usepackage{bm}
\begin{document}

\title{Optimal design of measurement network for neutronic activity field
reconstruction by data assimilation}

\author{Bertrand Bouriquet $^1$ \footnote{bertrand.bouriquet@cerfacs.fr}
   \and Jean-Philippe Argaud $^2$
   \and Romain Cugnart $^2$}

\maketitle

\footnotetext[1]{
Sciences de l'Univers au CERFACS, URA CERFACS/CNRS No~1875,
42 avenue Gaspard Coriolis,
F-31057 Toulouse Cedex 01 - France
}
\footnotetext[2]{
Electricit\'e de France,
1 avenue du G\'en\'eral de Gaulle,
F-92141 Clamart Cedex - France
}

\begin{abstract}

Using data assimilation framework, to merge information from model and
measurement, an optimal reconstruction of the neutronic activity field can be
determined for a nuclear reactor core. In this paper, we focus on solving the
inverse problem of determining an optimal repartition of the measuring
instruments within the core, to get the best possible results from the data
assimilation reconstruction procedure. The position optimisation is realised
using Simulated Annealing algorithm, based on the Metropolis-Hastings one.
Moreover, in order to address the optimisation computing challenge, algebraic
improvements of data assimilation have been developed and are presented here.

{\bf keyword:}  

Data assimilation, neutronic, activities reconstruction, nuclear in-core
measurements, inverse problem, network design, simulated annealing,
Metropolis-Hastings

\end{abstract}

\section{Introduction\label{sec:int}} 

Data assimilation methodology allows to build optimal reconstruction of activity
field, within a nuclear core, using information coming from both model and
measurements \cite{Parrish92,Todling94,Ide97}. The efficiency of data
assimilation for physical field reconstruction has already been demonstrated in
several articles in meteorology \cite{era40,Kalnay96,Huffman97}. Demonstration
of its efficiency for nuclear core activity field reconstruction was also done
in recent articles such as \cite{Massart07,Bouriquet2010,Bouriquet2011}. The
main points of data assimilation are presented in appendix \ref{sec:da}.

However, in all those applications of data assimilation, the measurement network
is considered to be known. The inverse problem consists then in optimizing the
location of the measuring instruments, in order to get the best possible results
from the data assimilation reconstruction procedure. Unlike the meteorological
domain, for which the question cannot be addressed due to the great size of the
problem (over $10^6$ measurement locations), the issue can be addressed and
resolved for the nuclear core, as only about $10^2$ to $10^3$ measurement
locations have to be taken into account.

The chosen method to optimise the instrument location is the well known
Simulated Annealing algorithm, based on a Metropolis-Hastings one
\cite{Metropolis53,Hastings70}. This algorithm was originally designed to deal
with problems of statistical physics. The first version, from N.~Metropolis 
\cite{Metropolis53}, was only focused on Boltzmann equation. Then W.~K.~Hastings
generalized it to other cases \cite{Hastings70}. From these methods,
S.~Kirkpatrick proposed some improvements \cite{Kirkpatrick83} by changing the
definition of the energy, used in the algorithm to optimise position in system
presenting an inner local order. Other applications of this algorithm already
exist for example in nuclear incident network design since a few years
\cite{Abida08,Abida09}. A brief description of the main points of Simulated
Annealing is presented in appendix \ref{sec:metro-methode}.

One of the key point in the Metropolis-Hastings algorithm is the permutation of
an instrumented location in the core with a non instrumented one. This can be
numerically costly to redo data assimilation calculations for each case, as it
includes big matrix inversions. To make those calculation feasible, we developed
methods of fast calculation, that treat instrument movement as a sequence of
loss and gain of instruments. This can be considered as a perturbation of the
initial state, which become then rather cheap to compute. The instrument loss
fast calculation was already used in \cite{Bouriquet2010}. The algebraic
acceleration method is recalled in appendix \ref{app:fastcalc}.

In the paper, the first section describes the general framework of the study. In
a second section, the parametrisation of simulated annealing method is provided.
Then the optimisation results, of simulated annealing coupled with an
assimilation procedure, are shown, either starting from standard reactor
instruments distribution or from a random distribution. In a next section, the
synthesis of both result is done. The global conclusion on the result of
instrument location optimisation using simultaneously simulated annealing and
data assimilation is then given.

\section{Definition of the nuclear core working framework\label{sec:def}}

\begin{figure}[!ht]
\begin{center}
  \includegraphics[width=7cm]{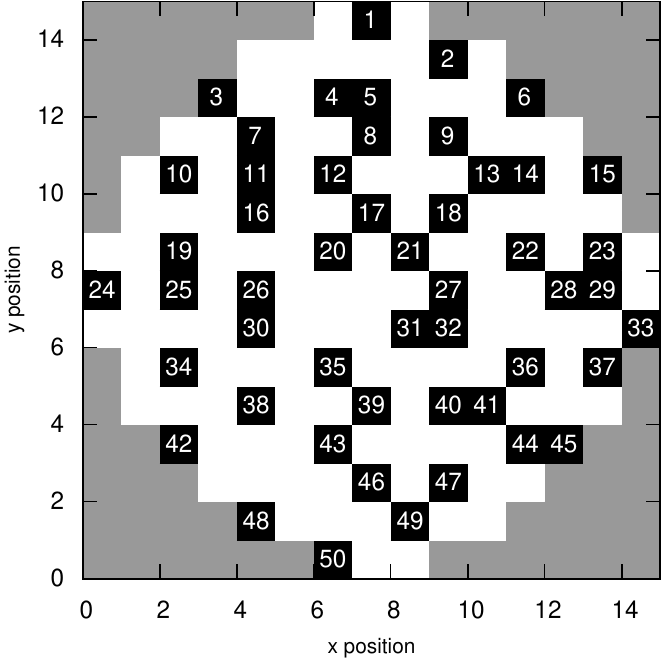}

  \caption{In the nuclear core, MFC instruments position is indicated by
  assemblies in black within an horizontal slice of the core. The assemblies
  without instrument are marked in white, and the neutronic reflector, out of
  the reactive core, is in grey. \label{fig:figMFC}}
\end{center}
\end{figure}

The experimental set up of this study is the standard configuration of a 900 MWe
nuclear Pressurized Water Reactor (PWR900), and we study the neutronic activity
field reconstruction. The core is filled with a total of $157$ vertical nuclear
fuel assemblies, among those $50$ are instrumented with Mobiles Fissions
Chambers (MFC) to measure the neutronic activity fields. An horizontal slice of
a PWR900 core is represented on the Figure \ref{fig:figMFC}. The question is
then to find the location of the $50$ instruments within the $157$ allowed
positions. We don't consider here any constraint on the location of
measurements, despite the fact that real constraints exists (mainly, MFC and
control rods, used for safety and steering, are incompatible). It is not
realistic, but, without these constraints, the optimisation problem is more
similar to experimental capabilities of new instruments on latest reactors, and
is far larger and difficult to solve. Solving this unconstrained optimisation
problem is then a prerequisite to take into account any constraint.

To perform data assimilation, both simulation code and experimentally measured
data are needed. For the simulation, the EDF newest experimental calculation
code COCAGNE for nuclear cores \cite{Courau2008,Hoareau2008} is used in a
standard configuration. A description of basic features of the underlying
neutronic model is done in the appendix \ref{sec:DAparam}. The physical
assemblies are numerically represented using $29$ vertical levels. Thus, the
size of the data assimilation control vector $\xx$ is $4553$ $(157 \times 29)$.
The size of the observation vector $\yo$ is $1450$ $(50 \times 29)$. All the
details of data assimilation parameters, derived from those choice of model and
observation space, are given in appendix \ref{sec:DAparam}. 

To have a good understanding of the instrumentation effect, we study various
scenario of instrument configurations (even some that do not exist operationally
and so cannot be tested experimentally). For that, synthetic data are used, that
allows to have an homogeneous approach all along the document and to use twin
experiment methodology. Synthetic data is generated from a model simulation,
filtered through an instrument model, and noised according to a predefined
measurement error density function (usually of Gaussian type). 

\section{Parametrisation of the simulated annealing algorithm} 

The global description of the well known simulated annealing and the
Metropolis-Hastings algorithms is done in the appendix
\ref{sec:metro-methode}. This section only focus on the specific parameters
related to the problem of instrument locations optimisation, coupled with a data
assimilation procedure for the activity field reconstruction. 

\subsection{Definition of the energy}

In the present case, the aim is not to minimise a real energy, but to improve
the quality of reconstruction $\xa$ of the neutronic activity fields on the
nuclear core. The reconstructed field $\xa$ is the result of a data assimilation
procedure, using parameters described in appendix \ref{sec:DAparam}, for a given
instrumental configuration.

Thus we would like to minimise the distance to the true value $\xt$ of the core
activity fields, that we know because we decide to set ourselves in twin
experiment framework. The energy $E$ will then be defined as the norm of the
difference between the analysis $\xa$ and the true value $\xt$, written as the
following formula:
\begin{equation}\label{Metro-Ext}
E(\xa) = || \xa - \xt||
\end{equation}

This definition of the energy will be used, all along the Metropolis-Hastings
algorithm, for the required term in formula \ref{eqboltz}.

\subsection{Measuring the evolution of the quality}

To measure the evolution of the quality of an instrument configuration, as a
function of the iteration in the process, a natural choice is to plot the value
of the energy $E(\xa)$.

However, this value as no real physical meaning. Thus, it is more interesting to
study the difference of $E(\xa)$ with respect to a chosen reference
$E(\xa_{ref})$. This is only a shift in the energy function, but this leads to a
better suited function to represent the quality of the $\xa$ reconstruction: the
best is the instruments location, the closest the new quality is to zero.

We choose this reference  value to be given by an analysis $\xa_{ref}$, obtained
by data assimilation, assuming each assembly in the core is instrumented with
one MFC. In a idealized case, such a "limit state" can be accessed by
experiment. The calculation is done with respect to the same parametrisation of
data assimilation as described in the appendix \ref{sec:DAparam}.

The quality evaluation factor $q$ is then given by the following formula:
\begin{equation}\label{quality}
q(\xa) = E(\xa) - E(\xa_{ref})  = || \xa - \xt|| - || \xa_{ref} - \xt||
\end{equation}
Its values are negative. Thus, if the Metropolis-Hastings algorithm improves the
instrument location choices, we expect to get an increasing curve for $q$ over
the iterations.

\subsection{Starting state for the optimisation algorithm}

The Metropolis-Hastings algorithm is an iterative algorithm, requiring a
starting point or instrument state to begin optimisation. We choose particular
starting points to illustrate how is working the algorithm. Moreover, the
Metropolis-Hastings algorithm is also a stochastic one. Thus assuming no seed is
chosen in the random number generator, two realisations of the algorithm will
lead to different sequences of evolution.

The classical PWR900 configuration is expected to be a very good one, reliable
and being designed years ago. Thus, as a first possibility for the starting
point of the Metropolis-Hastings algorithm, this configuration can be chosen.
This will assess the quality of this design (that originally satisfy also to a
lot of operational constraints, on the contrary to the one we are seeking by
optimization). In order to illustrate how is working the algorithm, two
different random realisations of the algorithm will be presented based on the
same initialisation in section \ref{sec:results1}. 

A second possibility for the starting point is to randomly choose it. This will
allow comparison of the optimal search with the one obtained with PWR900
configuration as initial point.

An another comparison can be done by using less instruments than the $50$
classically considered, in order to look at the influence of the number of
instruments. In this case, the starting point has to be randomly chosen, as
there is no standard repartition of instruments.

\subsection{Other parameters of the algorithm}

As is stated in appendix \ref{sec:metro-methode}, parameters need to be chosen
to run a Metropolis-Hastings algorithm. There are mainly $2$ parameters to set
in the present case.

The first one is the maximal number of iterations, which is the number of time
that we go through the loop on instrumentation swapping. This number will be set
to $1800$. This value was chosen for two reasons. The first point is that we
notice, through the setting up trials, that with such a number of iterations, we
reach a steady state in every case we processed. As a second point, it
corresponds to reasonable computing time of about $10$ hours (depending on the
computer used) per search and per processor.

The second parameter is the choice of the thermalisation function that fixes the
evolution of the pseudo-temperature in the algorithm. We experiment several
standard thermalisation formulation, and put our final choice on a classical
simple inverse function of the iteration, defined as follows:
\begin{equation} \label{equa-temp}
T_i = \frac{T_0}{i+1}
\end{equation}
where $i$ is the iteration step. The initial value $T_0$ of the temperature was
set to $0.05$. With such a definition, we got satisfactory result for the
optimisation algorithm.

\section{Result of optimisation from PWR900 configuration\label{sec:results1}}

\begin{figure}[!ht] 
\begin{center}
  \includegraphics[width=7cm]{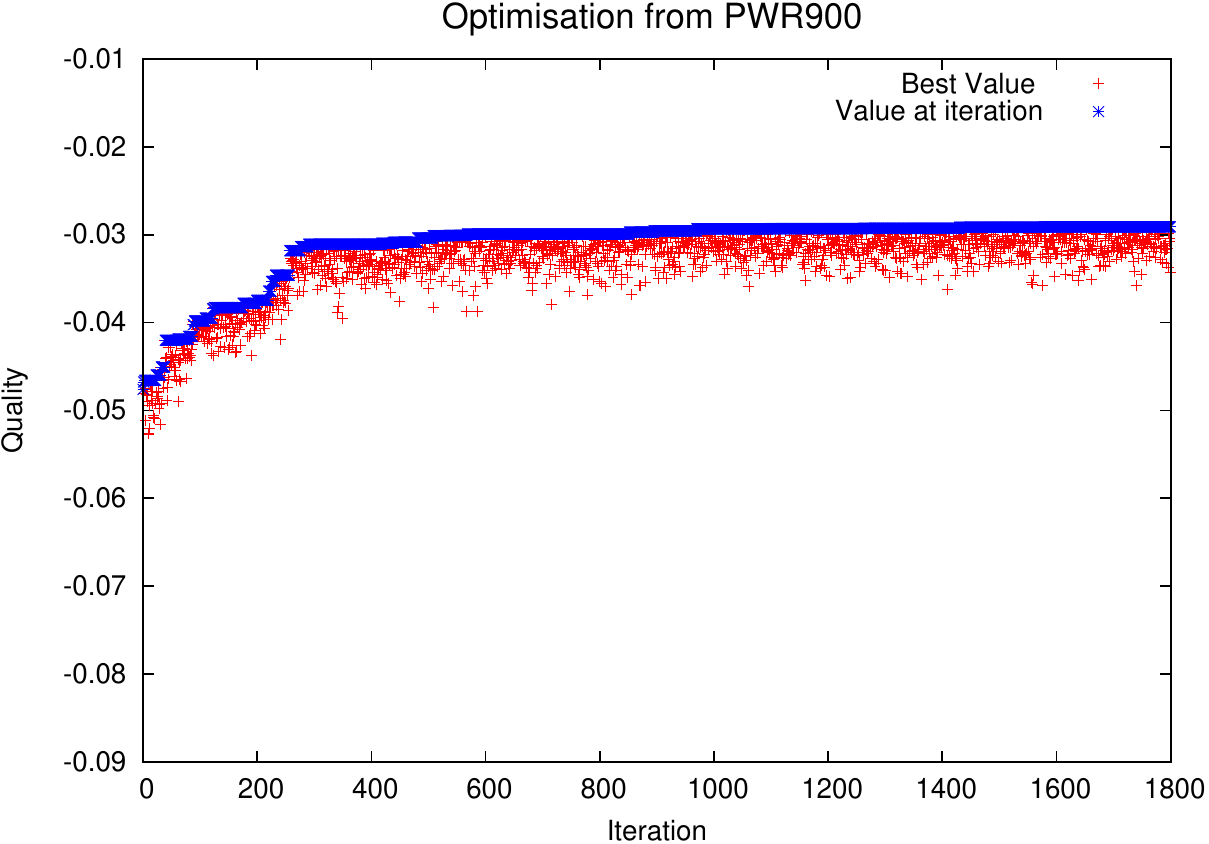} 
  \caption{Evolution of the quality of the reconstruction as a function of the
  iteration on 1800 steps (first random realization). At each step are
  represented the best evaluated quality as well as the quality at the current
  iteration.
  \label{fig:Metro-figx}}
  
  ~

  \includegraphics[width=7cm]{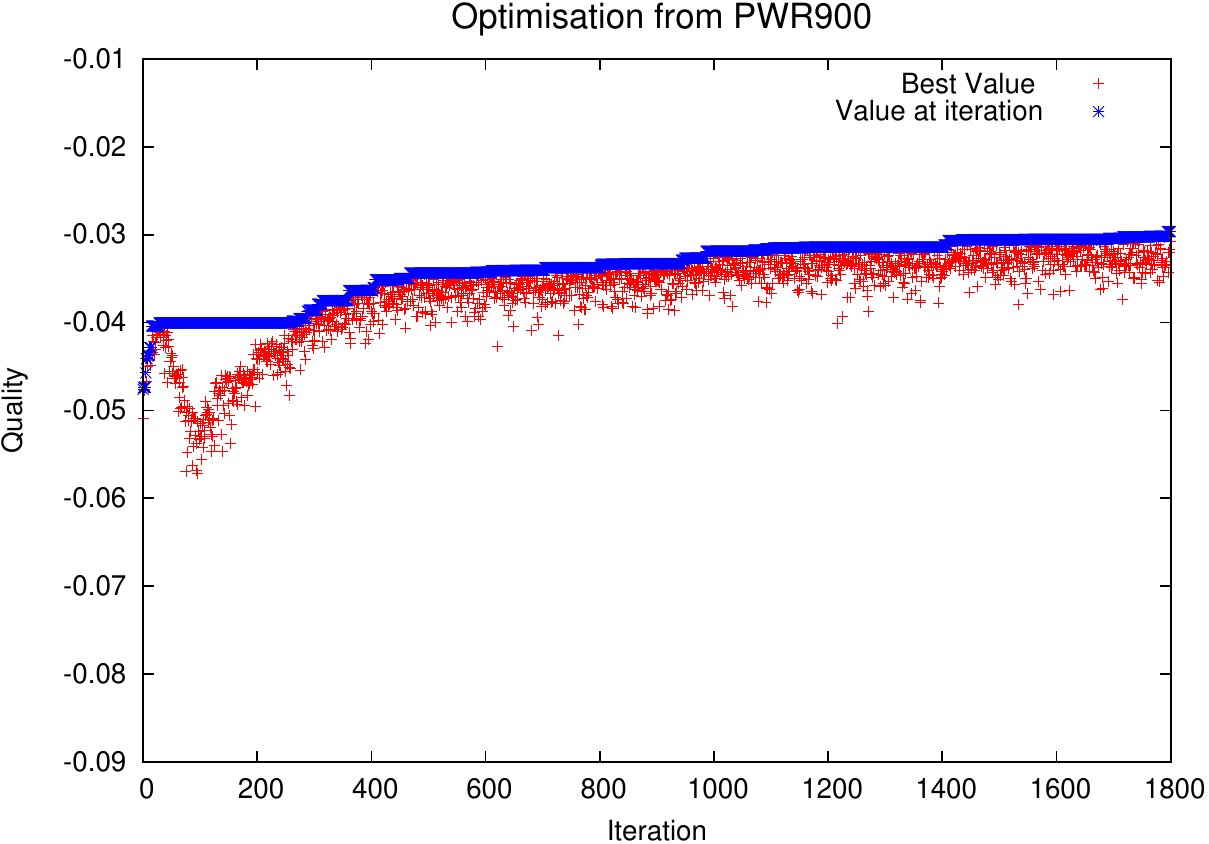}  
  \caption{Evolution of the quality of the reconstruction as a function of the
  iteration on 1800 steps (second random realization). At each step are
  represented the best evaluated quality as well as the quality at the current
  iteration.
  \label{fig:Metro-figx-r2} } 
\end{center}
\end{figure}

In the figure \ref{fig:Metro-figx} and \ref{fig:Metro-figx-r2}, two cases are
presented, based on the same initialisation and on different random realization.
On those figure, the red crosses represent the quality of the configuration at
the current iteration, and the blue asterisks represent the quality of the best
state founded since the beginning.

First, we are looking at the best state found curve. In both figures
\ref{fig:Metro-figx} and \ref{fig:Metro-figx-r2}, it can be noticed that the
algorithm reach a better state at the end than the initial one. Looking at the
quality of the final state in the two cases, it can be noticed that they are
very close. Looking on a set of realisation of the optimisation (not presented
here), the final state is always around the same level of quality with a rather
small dispersion (around $0.002$, that is around $5\%$). In both cases, in a
first phase, the quality of the configuration improves fairly quickly and a good
configuration can be found after around $300$ iterations. After that, the
improvement of the configuration is very slow and even stagnate above $800$
iterations.

Then, it is interesting to have a look at the value of the quality of the
current state for each iteration, to better understand the behaviour of the
Metropolis-Hastings algorithm. The plot of the current state in figure
\ref{fig:Metro-figx} is the typical one we got over several realisation. The
current state is "oscillating" around the current best state, to find a new
optimal state. The behaviour of the algorithm plotted in figure 
\ref{fig:Metro-figx-r2} is less common, but very illustrative of the
potentiality of the method. We notice that the quality of the final state is
about the same as the one of the figure \ref{fig:Metro-figx}. But in figure
\ref{fig:Metro-figx-r2}, we notice also that the current state can get very far
away of the best state, for a very long time steps. This kind of minimum search,
far from the current state, is one of the strong point of the present algorithm.
This research of a new optimal state is possible because of the highest pseudo
temperature at the beginning of the process given by equation \ref{equa-temp}.
The most interesting fact is that, whatever how far the method explore the state
space, finally it converge again to a state of good quality. This property of
Metropolis-Hastings algorithm, noticed qualitatively here, have been
demonstrated rather recently mathematically as mentioned in reference
\cite{Michel10}.

\begin{figure}[!ht] 
\begin{center}
  \includegraphics[width=7cm]{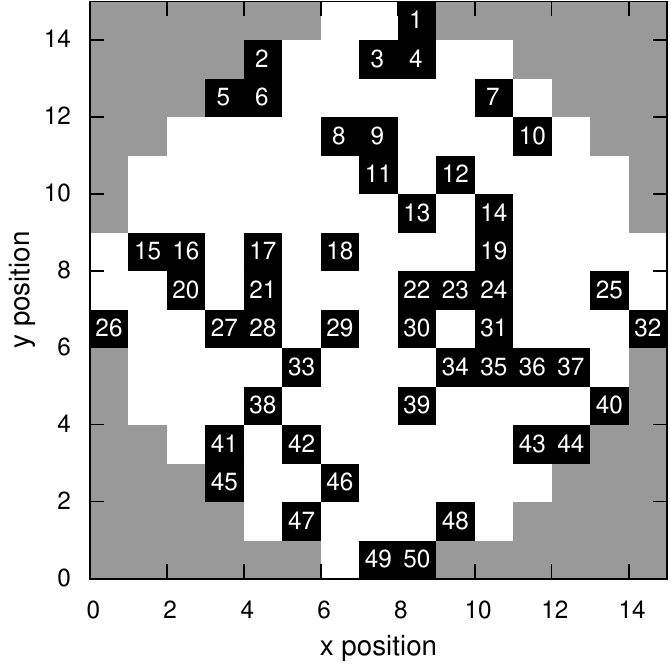} 
  \caption{The optimised positions of MFC instruments after a first random
  realisation are localised in assemblies in black, within an horizontal slice
  of the core. The assemblies without instrument are marked in white and the
  reflector, out of the reactive core, is marked in grey.
  \label{fig:Metro2D-figx}}
  
  ~

  \includegraphics[width=7cm]{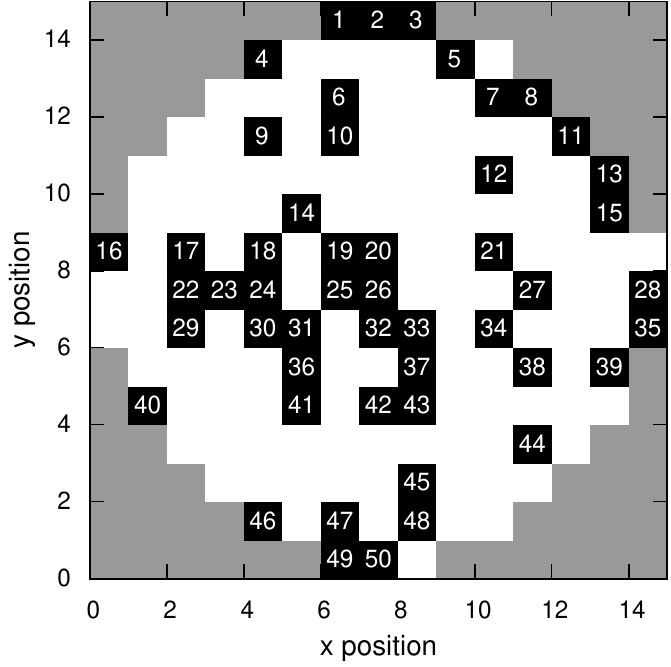}  
  \caption{The optimised positions of MFC instruments after a second random
  realisation are localised in assemblies in black, within an horizontal slice
  of the core. The assemblies without instrument are marked in white and the
  reflector, out of the reactive core, is marked in grey.
  \label{fig:Metro2D-figx-r2} } 
\end{center}
\end{figure}

At the end of the optimisation process, a new set of locations for the
instruments is obtained. The set of location coming from the two previous
optimisations are plotted in figures \ref{fig:Metro2D-figx} and
\ref{fig:Metro2D-figx-r2}.

Location set represented in \ref{fig:Metro2D-figx} and \ref{fig:Metro2D-figx-r2}
have, at glance, no common point. However, from the point of view of the quality
defined by equation \ref{quality}, those two repartitions are very close. This
prove not only that one optimal state does exist, but also that there is an
ensemble of optimal states of rather close quality.

Examining more precisely both figure \ref{fig:Metro2D-figx} and
\ref{fig:Metro2D-figx-r2} (and other realisation not presented here), some
common features can be discussed. In particular, there is a tendency to form
small clusters of instruments, mainly localised in the centre of core. Another
feature is the repartition of several individual instruments around the core.
However, a detailed study over many realisations shows that such features are
not significant.

\section{Result of optimisation from random configuration and dependency to
instruments number\label{sec:rand}}

Without an initial guess of the instrument repartition as the one of standard
PWR900, the natural choice of a starting repartition is to take them randomly. 
To be in the same condition as in the PWR900 case, we choose to use $50$
instruments randomly located in the core. Using random location is, at the same
time, a good opportunity to get information on method and result when less
instruments are available. Thus the case where only $10$ instruments are in the
core will be studied too.

To define the problem, a brief study of the property of the result of data
assimilation quality for random instruments repartition has been done. Over a
set of $1000$ random distribution of the $50$ instruments, it can be noticed
that the quality $q$ of all data assimilation reconstructions $\xa$ (with
respect to formula \ref{quality}) are distributed following as Gaussian
distribution of mean $m=-0.0832$ and a standard deviation $\sigma=0.0078$.

This already means that, with a value of $q=-0.0476$, the PWR900 standard
repartition is an excellent choice, with a far greater quality.

Assuming only $10$ instruments, the distribution of the quality $q$ of the
analysis $\xa$ still have Gaussian shapes but with a mean value of $m-0.0868$
and a standard deviation of $\sigma=0.076$. The very small difference between
the mean value of both distributions proves that, due to distance of influence
included in data assimilation procedure, inappropriate localisation of the
instruments can result in decreasing performance of the method, as already
noticed in reference \cite{Bouriquet2011}.

\begin{figure}[!ht] 
\begin{center}
  \includegraphics[width=7cm]{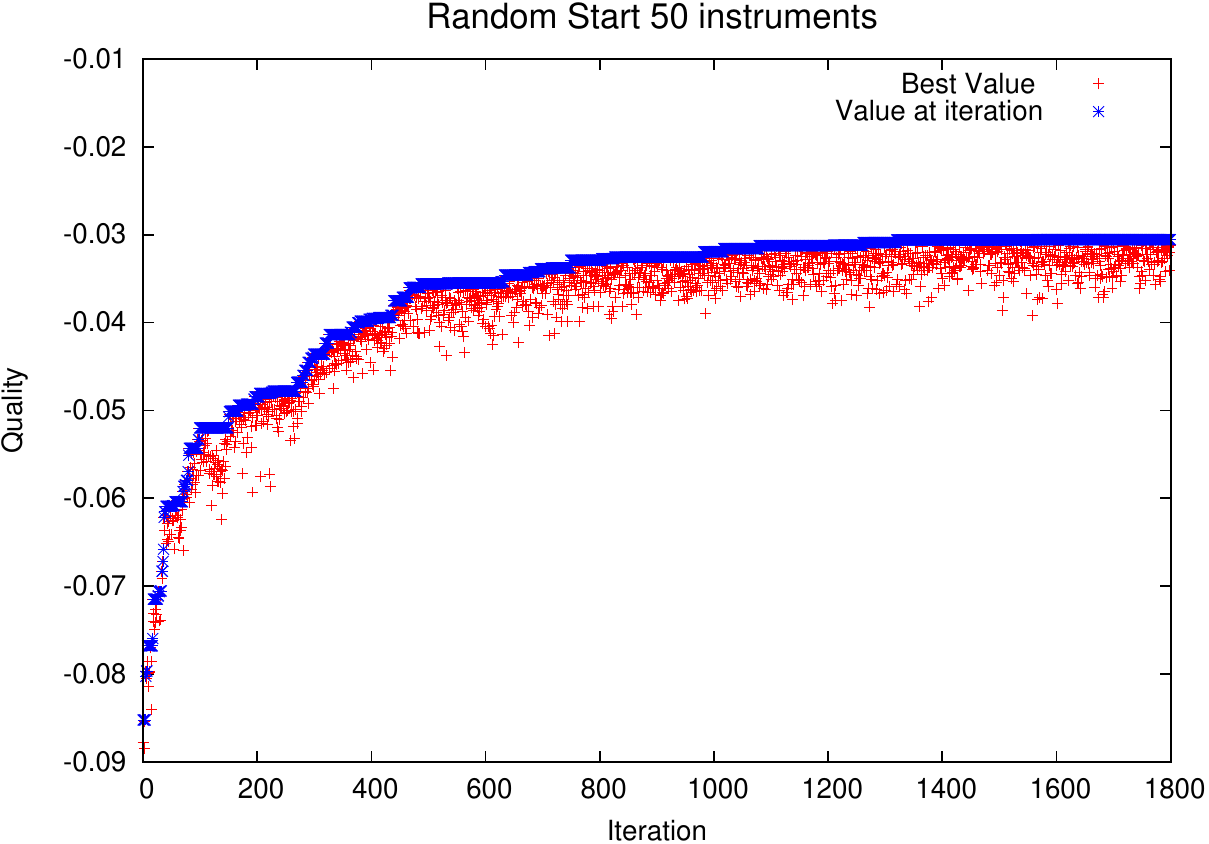} 
  \caption{Evolution of the quality of the reconstruction using $50$
  instruments, initially randomly localized, as a function of the iteration on
  1800 steps. At each step are represented the best evaluated quality, as well
  as the quality at the current iteration.
  \label{fig:Metro-figx-rand}}
  
  ~

  \includegraphics[width=7cm]{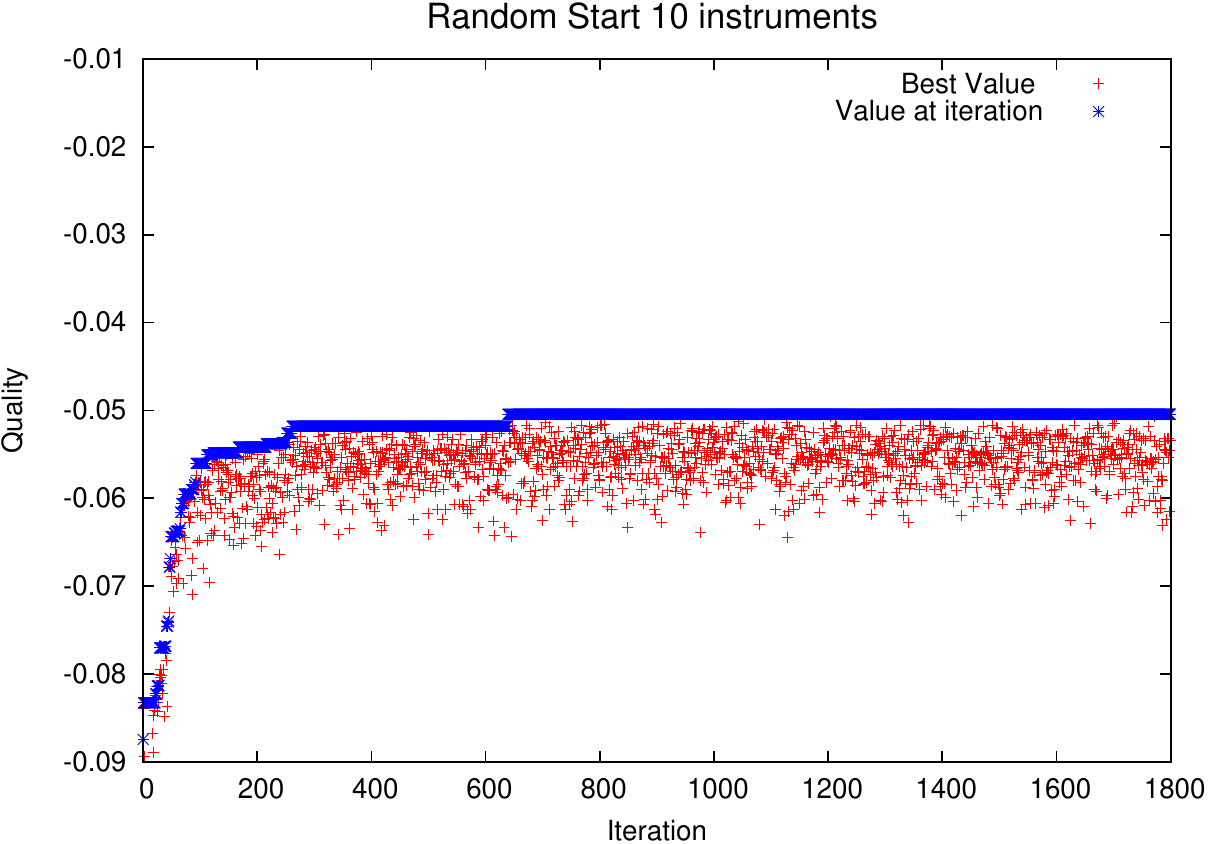}  
  \caption{Evolution of the quality of the reconstruction using $10$
  instruments, initially randomly localized, as a function of the iteration on
  1800 steps. At each step are represented the best evaluated quality, as well
  as the quality at the current iteration.
  \label{fig:Metro-figx-rand10} } 
\end{center}  
\end{figure}

Now we look at the evolution of the quality of the data assimilation
reconstruction as a function of iterations of Metropolis algorithm. In the
figures \ref{fig:Metro-figx-rand} and \ref{fig:Metro-figx-rand10}, those two
cases are presented.  Within those figures, as in figures
\ref{fig:Metro-figx} and \ref{fig:Metro-figx-r2}, the crosses in red
represent the quality of the configuration at the current iteration and the
asterisks in blue represent the quality of the best state founded.

In both figures \ref{fig:Metro-figx-rand} and \ref{fig:Metro-figx-rand10}, it
can be noticed a very fast improvement, of the quality of the instruments
setting, in very few iterations of the Metropolis-Hastings algorithm. This
initial phase is followed by a slow evolution of the quality, and then by a
stagnation phase. Stagnation phase is reached around after $600$ iterations
(resp. $300$) for $50$ (resp. $10$) instruments. This difference can be easily
explained, considering the space of possible configurations for $50$ instruments
in the core is far more large than the one of $10$ instruments. The final state
quality value is of $-0.0305$ (resp. $-0.0504$) for $50$ (resp. $10$)
instruments. Those qualities after optimisation are coherent with the amount of
available information, that is, of number of available instruments. This was
shown in \cite{Bouriquet2011}.

\begin{figure}[!ht] 
\begin{center}
  \includegraphics[width=7cm]{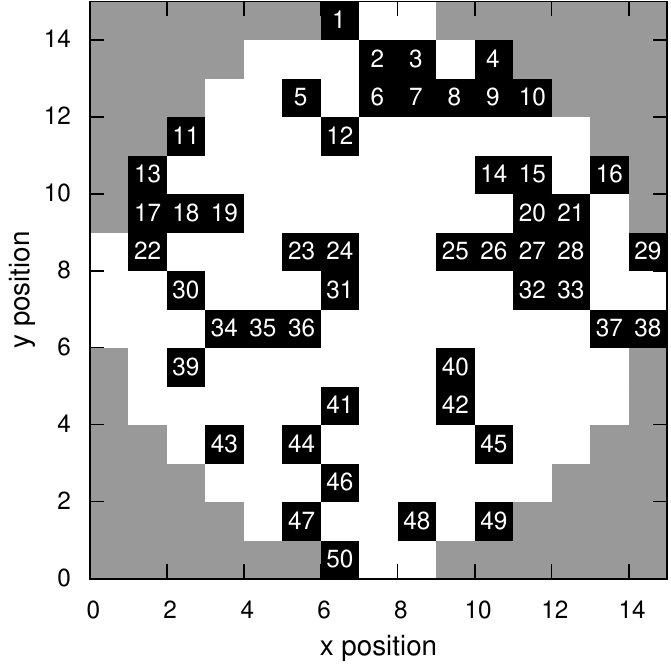} 
  \caption{The optimised locations of $50$ MFC instruments are localised in
  assemblies in black within the horizontal slice of the core. The assemblies
  without instrument are marked in white and the reflector, out of the reactive
  core, is in grey.
  \label{fig:Metro2D-figx-rand}}
  
  ~

  \includegraphics[width=7cm]{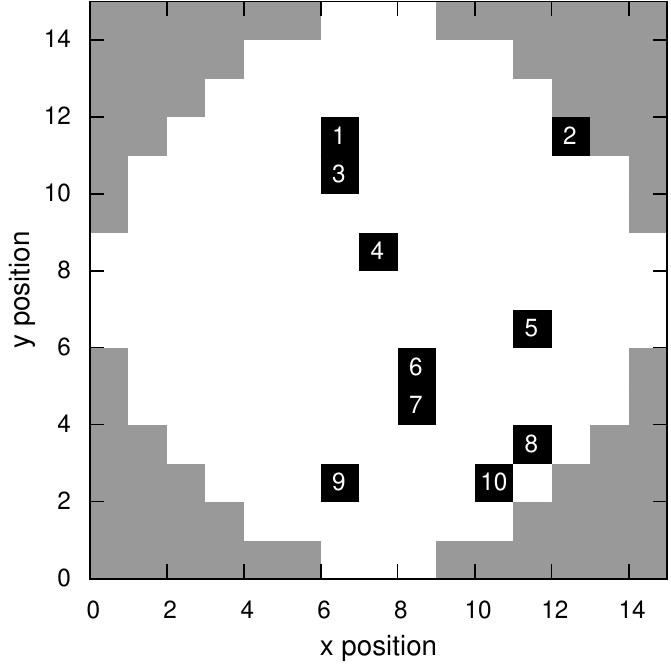}  
  \caption{The optimised locations of $10$ MFC instruments are localised in
  assemblies in black within the horizontal slice of the core. The assemblies
  without instrument are marked in white and the reflector, out of the reactive
  core, is in grey.
  \label{fig:Metro2D-figx-rand10} } 
\end{center}  
\end{figure}

The behaviour of the current state is typical of the Metropolis-Hastings
algorithm on both figures \ref{fig:Metro-figx-rand} and
\ref{fig:Metro-figx-rand10}. It is then interesting to compare the locations of
the instruments in the optimised state for both cases of $50$ and $10$
instruments. Those locations are presented in figures 
\ref{fig:Metro2D-figx-rand} and \ref{fig:Metro2D-figx-rand10}. 

In both figures \ref{fig:Metro2D-figx-rand} and \ref{fig:Metro2D-figx-rand10}
appear the same features as the one obtained in figures \ref{fig:Metro2D-figx}
and \ref{fig:Metro2D-figx-r2}, coming from optimisation of an initial PWR900
standard configuration. Those features are a tendency to form small cluster of
instruments mainly localised in the centre of core, and the repartition of
several instrument around the core. However, this is still statistically
irrelevant. 

\section{Synthesis of the results}

\begin{figure}[!ht]
\begin{center}
  \includegraphics[width=7cm]{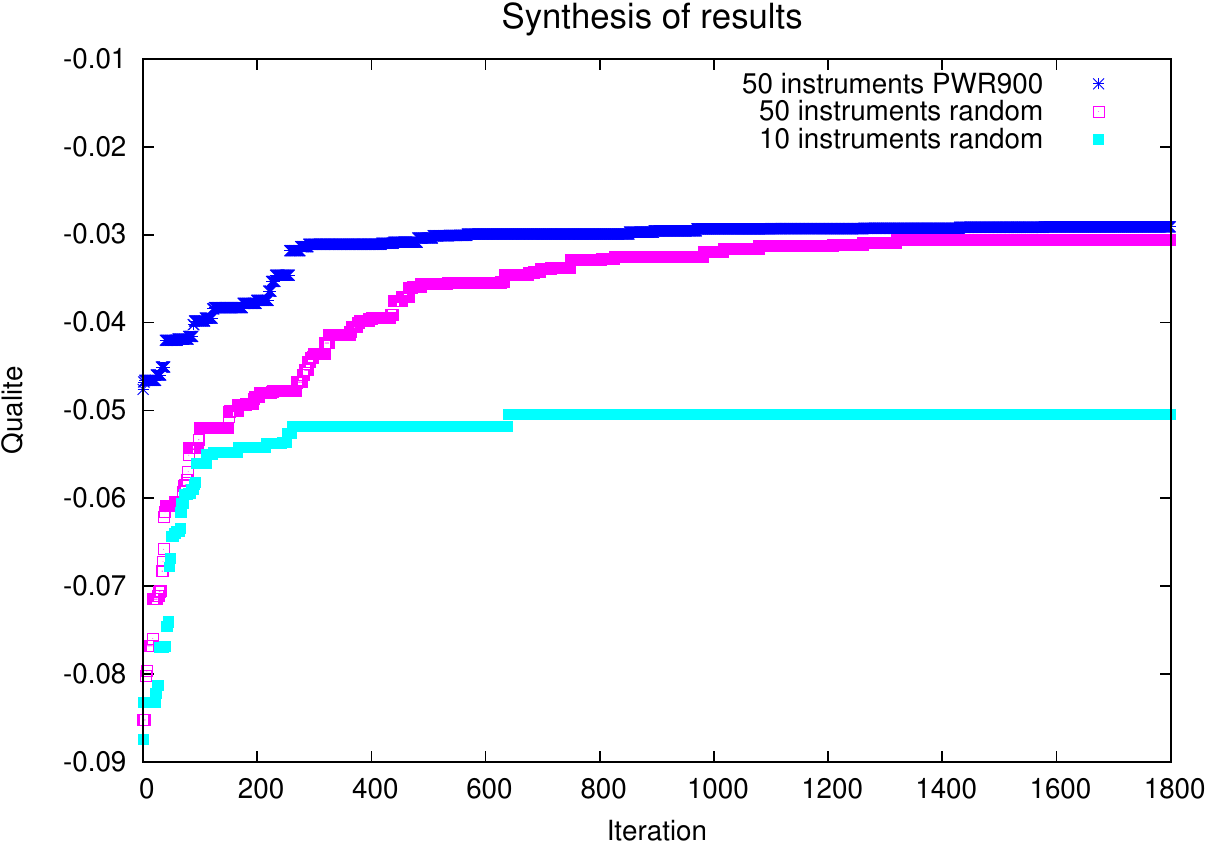}
  \caption{Evolution of the quality of reconstruction as a function of the
  iteration on 1800 steps, with $50$ instruments initialised by PWR900 or
  random configurations, and with $10$ instruments initialised randomly.
  \label{fig:Metro-figxt-compsynth}}
\end{center}
\end{figure}

Using the information gathered from the various uses of the method, we make a
synthetic plot of the information obtained and comment it. On figure 
\ref{fig:Metro-figxt-compsynth} are plotted the quality of the best evaluation
as a function of the number of iteration coming from figures
\ref{fig:Metro-figx}, \ref{fig:Metro-figx-rand} and
\ref{fig:Metro-figx-rand10}.

The first remark on figure \ref{fig:Metro-figxt-compsynth} is that, whatever are
the initial condition (instruments repartitions or number of instruments), the
simulated annealing method allows to find an better solution.

One interesting result of this study is, as expected, the good quality of the
PWR900 standard instruments distribution. Without any optimisation, the quality
of this repartition is $q=-0.0476$, that should be compared to a mean value of
$m=-0.0832$ for several random repartition with $50$ instruments. We recall that
the optimisation procedure is done without any constraints, which is not
physically comparable with the PWR900 and its standard instruments distribution.

Thus, this is not a good idea to chose randomly the position of the instruments,
even with a data assimilation procedure to make an optimal determination of the
activity field. However, looking for the case with $50$ instruments,  after 
optimisation, the final results have the same quality with a quality factor $q$
close to $-0.0325$. This is a experimental confirmation of the theoretical
result about convergence of simulated annealing method \cite{Michel10}, this
even if the final state for the same quality  do not look alike as it can be
seen in figures \ref{fig:Metro2D-figx} and \ref{fig:Metro2D-figx-rand}. 

Using different number of instruments as in section \ref{sec:rand}, we notice
interesting results. Observing the starting quality value $q$ of results with
data assimilation, the situation is rather blurry, as the quality with $50$
instruments is close to the one with $10$ instruments. Then, after position
optimisation by simulated annealing, ordering is clearly done according to the
instruments number: the more instruments we use, the best quality we get.
Moreover, this result is general, and final quality is always ordered respect to
the number of available instruments.

All those results demonstrate the efficiency and robustness of the simulated
annealing method for instruments position optimisation. This is even more
stringent considering that the method is applied in link with data assimilation
technique, that already enrich the global information on the whole system.

\section{Conclusion}

From this study, two important results are enlightened.

First, the standard PWR900 instruments repartition is characterized by an
excitingly good quality. This instruments repartition is {\it a priori} the best
we know, even if it was not originally designed using a data assimilation
framework background.

The second point is that the simulated annealing method can always find a
ameliorated instrument locations set. Moreover, as expected theoretically
\cite{Michel10}, the final state is of the same quality whatever is the starting
point. And using a reduced number of instruments, the final organisation has
always a quality that is increasing with the increase of the number of
instruments.

Those results demonstrate that, within the framework of neutronic simulation for
nuclear PWR, and using a reasonable size for the discrete space in which to
optimize the instruments (below $10^4$ points), it is possible to do
instruments repartition optimisation with data assimilation method. Such a work
can be done more efficiently considering an algebraic optimisation of the
calculations, and become still very reasonable in computing time.

Within the framework of nuclear core neutronic simulation, several developments
can be done from those results. Specially, instruments repartition can be done
under constraints such as symmetry or specific position for example. Moreover,
use of data assimilation open the way to obtain simulated annealing optimisation
with heterogeneous available instruments.

\bibliography{bibliographie}
\bibliographystyle{elsarticle-num}

\pagebreak
\appendix

\section{Metropolis-Hastings algorithm\label{sec:metro-methode}}

The algorithm used for position optimisation is the Metropolis-Hastings one. It
was initially used to describe the thermodynamically evolution of a system, and
the first algorithm was written by N. Metropolis, A. W. Rosenbluth,
M. N. Rosenbluth, A. H. Teller, and E. Teller
\cite{Metropolis53}. In this version, only a Boltzmann type distribution was
considered as it is the main one used in statistical physic. Then, in 1970, W.
K. Hastings extends this algorithm to other distributions \cite{Hastings70}.

The method is a random walk in instruments localisation space, and so an
iterative algorithm. This random walk use a probability $Q$ to chose the next
possible position $x$, moving from $x_k$ at step $k$. The transition probability
is $Q(x_k,x)$. Thus the probability that the whole system go from $x_k$ to $x$
at step $k+1$  (which means $x_{k+1}=x$) is then defined as follow:

\begin{equation} \label{eqn:base-metro}
P(x_{t+1} = x | x_t) = \min \left \{ \frac{\pi(x) Q(x_t,x)}{\pi(x_t) Q(x,x_t)}
,1   \right \}
\end{equation}

The remaining probability is the one to stay in $x_k$:

\begin{equation} 
P(x_{t + 1} = x_t | x_t) = 1 - P(x_{t + 1} = x | x_t).
\end{equation}

In the very common case of symmetric evolution, which means that  $Q(x,y) =
Q(y,x)$, some extra simplifications can be obtained. When optimising instruments
location, we are in such a case, as changing position of an instrument and then
making the invert operation leads to the same first repartition. Moreover, the
choice of both, the instrument to move, and its new localisation, are perfectly
random:
\begin{itemize}
\item with a probability of  $1$ if $\pi(x) \geq \pi(x_t)$
\item with a probability of $ \frac{\pi(x)}{\pi(x_t)}$ else 
\end{itemize}
using a given probability distribution $\pi$. Any kind of probability
distribution can be taken into account. In the present case, like in many
applications of Metropolis algorithm, a Boltzmann distribution will be used as 
probability function. The Boltzmann distribution is defined by the following
formula:
\begin{equation} 
\pi(x) = e^{\frac{-E(x)}{T}}
\end{equation}
where $E(x)=E$ is an evaluation of the quality of the state (such as a physical
energy or something else) of the state $x$, and where T is a temperature or a
pseudo-temperature associated to $E$. Thus, in the equation
\ref{eqn:base-metro}, when we make the ratio $\frac{\pi(x)}{\pi(x_t)}$, we
obtain the following formula:
\begin{equation} \label{eqboltz}
\pi(x) = e^{\frac{-E(x) + E(x_k)}{T}} = e^{\frac{-\Delta(x,x_k}{T}}
\end{equation}
This is this formula used here for the Metropolis algorithm.

The final algorithm of positions optimisation is then a loop, that last until a
condition on number of iteration or on energy limit is reached. Within the loop,
the following actions are realised:

\begin{itemize}
\item Swap two instruments or displace one instrument by swapping with a void
place,
\item Evaluate the pseudo energy $E$ of the new configuration,
\item Compare and record the best configuration founded,
\item Chose evolution of the instrument pattern according to probability law
given by equation \ref{eqboltz},
\item Go back to first step.
\end{itemize}

The algorithm is finally simple and reliable. Those type of methods have been
used for positions optimisation in several applications, and one of the first
and most relevant result was obtained for optimisation of chip design by  S.
Kirkpatrick, C.~D. Gelatt and M.~P. Vecchi \cite{Kirkpatrick83}.

\section{Data assimilation\label{sec:da}}

We briefly introduce the useful data assimilation key points to understand their
use, as applied in \cite{Talagrand97,Kalnay03,Bouttier99}.

Data assimilation is a wide domain and these techniques are, for example, the
keys of the nowadays meteorological operational forecasts \cite{Rabier2000}.
This is through advanced data assimilation methods that weather forecasting has
been drastically improved during the last 30 years. All the available data, from
satellites, aircraft or ground measurements, are used in conjunction with
complex numerical weather models.

The goal of data assimilation methods is to estimate the true value $\xt$ of the
state of the considered system, where the $t$ index stands for "\textit{true}".
The basic idea is to combine information from an \textit{a priori} knowledge on
the state of the system (usually denoted as $\xb$, with $b$ for
"\textit{background}"), and some measurements (referenced as $\yo$, with $o$ for
"\textit{observations}"). The background is usually the result of numerical
simulations, but can also be derived from any {\it a priori} knowledge. The
result of data assimilation is called the analysis, denoted by $\xa$, and it is
an estimation of the true state $\xt$ we look for.   

The control and observation spaces are not necessarily the same, and a bridge
between them has to be built. This is the observation operator $H$, that
transforms values from the space of the background to the space of observations.
For data assimilation purpose, we use the linearised operator $\oH$ of $H$
around the background $\xb$. The reverse operator, converting observation
increments to background increments, is given by the transpose $\oH^T$ operator
of $\oH$.

Two other ingredients are necessary. The first one is the covariance matrix
$\oR$ of observation errors, defined as $\oR=E[(\yo-H(\xt)).(\yo-H(\xt))^T]$,
where $E[.]$ is the mathematical expectation. It is obtained from the known
errors assuming unbiased measurements, which means $E[\yo-H(\xt)]=0$. The second
one is the covariance matrix $\oB$ of background errors, defined as
$\oB=E[(\xb-\xt).(\xb-\xt)^T]$. These errors on the \textit{a priori} state are
also assumed it to be unbiased. There are many ways to get this \textit{a
priori} state and background error matrices. However, those matrices are
commonly obtained from the output of a model by an evaluation of accuracy, or
are the result of expert knowledge. 

Within this formalism, under a static assumption of state equations, the
analysis $\xa$ is the Best Linear Unbiased Estimator (BLUE), and is given by the
following equation:

\begin{equation}\label{xa}
\xa = \xb + \oK \big(\yo- H\xb\big),
\end{equation}
where $\oK$ is the gain matrix such as:
\begin{equation} \label{Kmat}
\oK = \oB\oH^T (\oH\oB\oH^T + \oR)^{-1}.
\end{equation}
Moreover, we can express the analysis error covariance matrix $\oA$
characterising the analysis errors $\xa-\xt$. This matrix is derived from $\oK$
as:
\begin{equation}
\oA = (\oI - \oK\oH)\oB,
\end{equation}
where $\oI$ is the identity matrix.

It is worth noting that solving Equation \ref{xa} is, if the probability
distribution is Gaussian, equivalent to minimise the following function
$J(\xx)$, $\xa$ being the optimal solution:
\begin{equation}\label{J}
J(\xx) = (\xx-\xb)^T \oB^{-1}(\xx-\xb) \medskip + \big(\yo-\oH\xx\big)^T \oR^{-1} \big(\yo-\oH\xx\big).
\end{equation}

This minimisation is known in data assimilation as 3D-Var methodology
\cite{Talagrand97}.

\section{Data assimilation implementation \label{sec:DAparam}} 

\subsection{Brief description of the nuclear core modelling\label{model}}

The aim of a neutronic code, like COCAGNE \cite{Courau2008,Hoareau2008} from
EDF, is to evaluate the neutronic activity field and all associated values
within the nuclear core. This field depends on the position in the core and on
the neutron energy. To do such an evaluation, the population of neutrons are
divided in several groups of energy. Classically, two energy groups are
considered, describing the neutronic flux by $\Phi=(\Phi_1,\Phi_2)$. The
material properties depend on the position in the core. The neutronic flux
$\Phi$ is identified by solving two-group diffusion equations described by:
\begin{equation}\label{eq:cirep:eqn}
\left\{
\begin{array}{l}
\displaystyle - \mbox{div}(D_1\mbox{\textbf{grad}}\Phi_1) + (\Sigma_{a1} +
\Sigma_r ) \Phi_1 \\
\displaystyle \quad\quad\quad\quad\quad\quad =  \frac{1}{k} \Big(\nu_1
\Sigma_{f1} \Phi_1 + \nu_2 \Sigma_{f2} \Phi_2 \Big) \\
\displaystyle
-\mbox{div}(D_2\mbox{\textbf{grad}}\Phi_2) + \Sigma_{a2} \Phi_2 - \Sigma_r
\Phi_1 = 0
\end{array}
\right.
\end{equation}
where $k$ is the effective neutron multiplication factor, all the quantities and
the derivatives (except $k$) depend on the position in the core, $1$ and $2$ are
the energy group indexes, $\Sigma_r$ is the scattering cross section from group
$1$ to group $2$, and for each group, $\Sigma_a$ is the absorption cross
section, $D$ is the diffusion coefficient, and $\nu\Sigma_f$ is the corrected
fission cross section.

The cross sections also depend implicitly on the concentration of boron, which
is a substance added in the water used for the primary circuit to control the
neutronic fission reaction. This control is described through a feedback
supplementary model. This model takes into account the temperature of the
materials and of the neutron moderator, given by external thermal and
thermo-hydraulic models. A detailed description of the core physic and numerical
solving can be found in reference \cite{Duderstatd76}.

The overall numerical resolution consists in searching for boron concentration
such that the eigenvalue $k$ is equal to $1$, which means that the neutron
production in the core is stable and self-sustaining. It is named
"\textit{critical boron}" concentration computation.

The activity in the core is obtained through a combination of the fluxes
$\Phi=(\Phi_1,\Phi_2)$. Numerically, the activity is given on a chosen mesh of
the core. Using homogeneous materials for each assembly (for example $157$ in a
classical EDF PWR900 reactor), and choosing a vertical mesh compatible with the
core (usually $29$ vertical levels), this result is an activity field of size
$157 \times 29=4553$ that cover all the core.

\subsection{The observation operator $H$}

The observation operator $H$ is, in the present application, a selection
procedure. This procedure extracts the values corresponding to effective
measurements among the values of the numerical model space. The normalisation
procedure is a scaling of the value with respect to the geometry and power of
the core. The entire process is linear. Then the linear operator $\oH$ is
identical to $H$. Size of the operator depend directly on the number of
instruments available. In a PWR900 as we treat it, the size is $(1450 \times
4553)$. 

\subsection{The background error covariance matrix $\oB$}

The $\oB$ matrix represents the covariance between the spatial errors for the
background. In order to get those, we estimate them as the product of a
correlation matrix $\oC$ by a normalisation factor.  

The correlation $\oC$ matrix is built using a positive function that defines the
correlations between instruments with respect to a pseudo-distance in model
space. Positive functions have the property (via Bochner theorem) to build a
symmetric defined positive matrix when they are used as matrix generator
\cite{Matheron70,Marcotte08}. In the present case, Second Order Auto-Regressive
(SOAR) function is used to prescribe the $\oC$ matrix. In such a function, the
amount of correlation depends from the Euclidean distance between spatial
points. The lengths of radial and vertical correlation (denoted as $L_r$ and
$L_z$ respectively, and associated to the radial $r$ coordinate and the vertical
$z$ coordinate) have different values, which means we are dealing with a global
pseudo euclidean distance. The function can be expressed as follow:
\begin{equation}  
C(r,z) = \left(1+\frac{r}{L_r}\right) \left(1+\frac{|z|}{L_z}\right)
         \exp{\left(-\frac{r}{L_r}-\frac{|z|}{L_z}\right)}.
\label{eqB}
\end{equation}
The matrix obtained by the above Equation \ref{eqB} is a correlation matrix. It
is then multiplied by a suitable variance coefficient to get a covariance
matrix. This coefficient is obtained by statistical study of differences between
model and measurements on real case. In our case, the size of the  $\oB$  matrix
is related to the size of model space, so it is $(4553 \times 4553)$.

\subsection{The observation error covariance matrix $\oR$}

The observation error covariance matrix $\oR$ is approximated by a diagonal
matrix. This assumes that no significant correlation exists between the
measurement errors of the measure instruments. On the diagonal, the usual
modelling is to take each value as a percentage of the corresponding observation.
This can be expressed as:
\begin{equation}  
\oR_{jj} = \left( \alpha (y^o)_j \right) ^{2}, \quad \forall j
\label{eqR}
\end{equation}
The parameter $\alpha$ is fixed according to the accuracy of the measurement and
of the representative error associated to the instrument. The size of the  $\oR$
matrix is related to the size of observation space, so it is here of $(1450
\times 1450)$ for a PWR900.

\section{Fast calculation method for data assimilation on perturbed instruments
network \label{app:fastcalc}}

The direct calculation of influence of the displacement of an instrument, which
is mandatory for the Metropolis algorithm, is very time consuming because of the
size of the matrices. Thus, to shorten this computing time, we developed special
methods to calculate the removal or the addition of an instrument as a
perturbation of a known state. Thus, changing the position of an instruments in
the network corresponds to remove an instrument and then to add one, or
inversely. Calculating the new analysis using such approach is then
substantially faster. We present in this section the two methods, for fast
analysis calculation when adding or removing instrument. Both techniques are
based on Schur complement \cite{Zhang05}.

\subsection{Instrument removal \label{appsec:remov}}

Within the BLUE assimilation method, the limiting factor in calculation time are
the matrices inversions. In Equation \ref{Kmat}, the costly part is the
inversion of the term $\oM$ defined as:
\begin{equation} \label{Mmat}
  \oM = \oH \oB \oH^T +\oR.
\end{equation}
In the present case, the number of terms of $\oM$ is around $2.10^6$, so 
the inversion at each step is rather time consuming when iterating for the
Metropolis algorithm. Then we try to optimise the computing cost. 

We noticed that the calculations are more time consuming when only few
instruments are removed. In this case the $\oM$ matrix remains still huge. 

Thus, the idea is to use the information obtained in the inversion of the full
size matrix to shorten calculation, by calculating smaller size matrix. In this
case, we want to calculate the new matrix as a perturbation of the original one.
This is done by exploiting the Schur complement of the matrix. 

We want to suppress some instruments to a given physical configuration. With
respect to the Equation \ref{Kmat}, we need to calculate a new matrix $\oK_n$.
The $n$ index is standing to designate the new matrix we want to calculate.
According to Equation \ref{Mmat}, we have to determine a new matrix $\oM_n$. 

This matrix $\oM_n$ is obtained from the knowledge of the invert of the matrix
$\oM_g$ calculated over all the instruments. The subscript $g$ is used to
designate the reference matrix we start from, according to Equation \ref{Mmat},
in the case of a complete initial instruments repartition.

All the components of the new matrix $\oK_n$ can be obtained by suppressing the
lines and columns corresponding to removed instruments in $\oM_g$, inverting it
and then multiplying this matrix by the corresponding $\oH_n$ and $\oB_n$. We
notice that, in our case, $\oB_n=\oB_g$ as procedure do not affect the model
space.

To make the explanation easier (and without loosing generality), we assume that
the suppressed instruments correspond to the lower square of $\oM_g$. If it is
not the case, it is always possible to reorganise the matrix in such a way. 

Now we put the matrix $\oM_g$ under a convenient form, separating remaining
measures from removed ones. Assuming the starting matrix $\oM_g$ is of size $m
\times m$, and assuming we plan to keep $p$ measurements and to suppress $s$,
$\oM_g$ can be written in the following way:
\begin{equation}   \label{pgi:dec}
\oM_g =
\left(
\begin{array}{cc} 
\oP_g & \oQ_g  \\ 
\oR_g & \oS_g  
\end{array} 
\right),
\end{equation}
where:
\begin{itemize} 

\item $\oP_g$ contains the remaining measurements, and is a $p \times p$ matrix,

\item $\oS_g$ contains the suppressed measurement, and is a $s \times s$ matrix,

\item$\oQ_g$ et $\oR_g$ represent the dependence between remaining measured and
suppressed ones. In the particular case we are dealing with, $\oQ_g^T=\oR_g$.
However, no further use of this property is done.

\end{itemize}
With such a decomposition, we got the equality $m=p+s$. The $\oP_g$ matrix
corresponds to the remaining instruments, thus we have the equality:
\begin{equation} \label{pgi:Pmat}
\oP_g  = \oM_n,
\end{equation}
The Equation \ref{pgi:dec} gives the decomposition  required to build the Schur
complement of this matrix \cite{Zhang05}. Under the condition that $\oP_g$ can
be inverted, the Schur complement is the following quantity:
\begin{equation} \label{pgi:shur}
\oS_g - \oR_g \oP_g^{-1} \oQ_g,
\end{equation} 
and is noted $(\oM_g/\oP_g)$. This notation reads as "\textit{Schur complement
of $\oM_g$ by $\oP_g$}".

Thus we look for a cheap way to calculate $\oP_g^{-1}$ knowing $\oM_g^{-1}$. For
that, we use the Banachiewicz formula \cite{Zhang05}, that gives invert of
$\oM_g$ as a function of $\oP_g$, $\oQ_g$, $\oR_g$, $\oS_g$ and $(\oM_g/\oP_g)$
matrices:
\begin{eqnarray}  \label{pgi:banach}
\oM_g^{-1} & = &
\left (
\begin{array}{cc} 
\oP_g & \oQ_g  \\ 
\oR_g & \oS_g  
\end{array} 
\right )^{-1}  
\\
 & = & \left (
\begin{array}{cc} 
\oP_g^{-1} + \oP_g^{-1} \oQ_g (\oM_g/\oP_g)^{-1} \oR_g \oP_g^{-1} & -\oP_g^{-1} \oQ_g (\oM_g/\oP_g)^{-1} \\ 
-(\oM_g/\oP_g)^{-1} \oR_g \oP_g^{-1} & (\oM_g/\oP_g)^{-1}
\end{array} 
\right ) \nonumber
\end{eqnarray}  
We define the 4 sub-matrices $\tP_g$, $\tQ_g$, $\tR_g$ and $\tS_g$ by:
\begin{eqnarray} 
\tP_g & = & \oP_g^{-1} + \oP_g^{-1} \oQ_g (\oM_g/\oP_g)^{-1} \oR_g\oP_g^{-1},\\
\tQ_g & = & -\oP_g^{-1} \oQ_g (\oM_g/\oP_g)^{-1},\\
\tR_g & = & -(\oM_g/\oP_g)^{-1} \oR_g \oP_g^{-1},\\
\tS_g & = & (\oM_g/\oP_g)^{-1}.
\end{eqnarray} 
Rearranging those terms we get:
\begin{equation} 
\oP_g^{-1} = \tP_g - \tQ_g \tS_g^{-1} \tR_g.
\end{equation} 
By hypothesis, the inverse $\oM_g^{-1}$ of the global matrix is known, and it is
possible to extract $\tP_g$, $\tQ_g$, $\tR_g$ and $\tS_g$ from the whole
inverted matrix. Thus, the main cost to obtain the inverse of $\oP_g$, of size
$p \times p$, becomes the inversion of the matrix $\tS_g$, which size is $q
\times q$. If the number of measurements to suppress is smaller than the number
of remaining ones, this methods gives a notable gain. As soon as the matrix
$\oP_g^{-1}=\oM_n^{-1}$, final calculation of $\oK_n$ is straightforward.

\subsection{Fast calculation of the gain matrix $\oK$ in the case of instruments
gain \label{appsec:gain}}  

The next step is to calculate the required matrix in the case of instruments
gain. In this case the situation is more tricky, as the dependency between the
various instruments need to be taken into account, if they do exist. As
previously, the technique used is based on Schur complement.  

In this part, we consider a system with two instruments labelled with number $1$
and $2$, and characterised by observation operators $\oH_1$ et $\oH_2$. We note
$d_1$ and $d_2$ the size of the spaces respectively associated with $\oH_1$ and
$\oH_2$. The size of the data space is denoted $n$. The matrix $\oH_1$ and
$\oH_2$ are of size $d_1 \times n$ and $d_2 \times n$ respectively.

The two error covariance matrices on measurement $\oR_1$ and $\oR_2$ are of
size  $d_1 \times d_1$ and  $d_2 \times d_2$ respectively. Without loosing
generality, the $\oB$ matrix can be assumed to be unique. Both operators 
$\oH_1$ and $\oH_2$ are application from the same space of data toward different
observation spaces. The dimension of the observation space is $n$, then the size
of this matrix $\oB$ is $n \times n$.

Starting from the formula \ref{Kmat} given for independent calculation of the
analysis, we got :
\begin{equation}\label{pgi:lossbasis}
\left\{
\begin{array}{cc} 
\oK_1 = \oB \oH_1^T (\oH_1 \oB \oH_1^T + \oR_1)^{-1} \\
\\ 
\oK_2 = \oB \oH_2^T (\oH_2 \oB \oH_2^T + \oR_2)^{-1}
\end{array}
\right.
\end{equation}  

The coupled observation operator of the whole measurement system is defined by
combination of the instruments $1$ and $2$. Thus, the complete observation
operator can be written as follow:
\begin{equation}  
\oH = \left( \begin{array}{c}
  \oH_1 \\ 
  \oH_2
\end{array} \right)
\end{equation}  
This is completely equivalent to exchange the two observation operators $\oH_1$
and $\oH_2$. The final matrix obtained is of size $(d_1 + d_2) \times n$.
Looking now at the structure of the covariance error matrix $\oR$, it can be
written as follow:
\begin{equation} 
\oR =
\left( \begin{array}{cc} 
\oR_1      & \oR_{12}  \\ 
\oR_{12}^T & \oR_2  
\end{array} \right)
\end{equation}  
In the above formula, the $\oR_{12}$ sub-matrix is the correlation error matrix
between the two instruments set. This sub-matrix contains the information
required to combine both assimilation. Even if we restart a data assimilation
from the beginning, this information need to be added. Without any information,
in many case, it can be assumed that no correlation does exist between the
errors of each set of instruments. This is equivalent to take $\oR_{12}=0$.
However, to keep generality, we will treat the problem assuming a non zero
$\oR_{12}$ 
matrix.

From that point, all the mandatory element to realise data assimilation
calculation are available. The complete matrix $\oK$, described in Equation
\ref{Kmat}, can be then decomposed according to the above information as follow:
\begin{equation}
\begin{array}{c}
\oK = \oB
\left(
\begin{array}{cc} 
\oH_1^T  & \oH_2^T \\
\end{array}
\right) \times 
\\
\left(
  \left( \begin{array}{cc} 
    \oR_1 & \oR_{12}  \\ 
    \oR_{12}^T & \oR_2  
  \end{array} \right)
  +
  \left( \begin{array}{c} 
    \oH_1  \\ 
    \oH_2  
  \end{array} \right)
  \oB
  \left( \begin{array}{cc} 
    \oH_1^T  & \oH_2^T \\
  \end{array} \right)
\right)^{-1}
\end{array}
\end{equation}  
Doing the multiplication of $\oH_1$ and $\oH_2$ matrix with $\oB$ matrix, the
previous expression becomes:
\begin{equation}\label{eqn:Kdecomp}
\begin{array}{c}
\oK =\oB
\left( \begin{array}{cc} 
  \oH_1^T  & \oH_2^T \\
\end{array} \right) \times 
\\
\left(
  \left( \begin{array}{cc} 
    \oR_1 & \oR_{12}  \\ 
    \oR_{12}^T & \oR_2  
  \end{array} \right)
  +
  \left( \begin{array}{cc} 
    \oH_1 \oB \oH_1^T & \oH_1 \oB \oH_2^T \\ 
    \oH_2 \oB \oH_1^T & \oH_2 \oB \oH_2^T
  \end{array} \right)
\right)^{-1}
\end{array}
\end{equation}  
In the term that needs to be inverted, the matrix is decomposed as a sum of
diagonal element and extra diagonal ones. The diagonals terms correspond to
previously known elements, and the extra diagonal ones appear due to the
interdependency between variables, both in the model (with terms
$\oH_1\oB\oH_2^T$ and $\oH_2 \oB \oH_1^T$) and in the observation space (with
term $\oR_{12}$). Thus, the equation \ref{eqn:Kdecomp} can be rewritten as :
\begin{equation}\label{pgi:Kdev}
\begin{array}{c}
\oK = \oB
\left( \begin{array}{cc} 
  \oH_1^T  & \oH_2^T \\
\end{array} \right) \times 
\\
\left(
  \left( \begin{array}{cc} 
    \oR_1+ \oH_1 \oB \oH_1^T & \Zero  \\ 
    \Zero & \oR_2 + \oH_2 \oB \oH_2^T
  \end{array}\right)
\right.
+ \\
\left.
  \left( \begin{array}{cc} 
    \Zero & \oH_1 \oB \oH_2^T + \oR_{12} \\ 
    \oH_2 \oB \oH_1^T + \oR_{12}^T & \Zero  
  \end{array} \right)
\right)^{-1}
\end{array} 
\end{equation}  
To simplify the notation, let:
\begin{equation} 
\begin{array}{cc} 
\oP_1 = \oH_1 \oB \oH_1^T + \oR_1 \\ 
\oP_2 = \oH_2 \oB \oH_2^T + \oR_2
\end{array} 
\end{equation}
and:
\begin{equation} 
\oP =
\left (
\begin{array}{cc} 
\oP_1 & \Zero  \\ 
\Zero & \oP_2
\end{array} 
\right )
\end{equation}  

If, as assumed, the data assimilation procedure for each of the instruments has
already been done, the terms $\oP_1$ and $\oP_2$ are available as well are their
inverse. Thus only extra diagonal term given by the following equation need to
be treated:
\begin{equation}\label{eqn:subterm}
\left(
\begin{array}{cc} 
\Zero & \oH_1 \oB \oH_2^T + \oR_{12} \\ 
\oH_2 \oB \oH_1^T + \oR_{12}^T & \Zero  
\end{array} 
\right )
\end{equation}  
Also to simplify the notation, we denote:
\begin{equation}
\oT = \oH_1 \oB \oH_2^T + \oR_{12}
\end{equation}  
The other term of the matrix of Equation \ref{eqn:subterm} can then be obtained
by transposing the previous $\oS$ matrix:
\begin{equation}
\oT^T  = \oH_2 \oB \oH_1^T + \oR_{12}^T
\end{equation}
It worth to recall that $\oB$ is a symmetric positive defined matrix, thus the
equality $\oB^T=\oB$ can be used.

Now, the aim is to put the part to invert in equation \ref{pgi:Kdev} under a
specific form, that will ease the inversion. In this purpose, we write:
\begin{equation} 
\oU =
\left (
\begin{array}{cc} 
\Zero & \oT  \\ 
\oI & \Zero  
\end{array} 
\right )
\end{equation}
and:
\begin{equation} 
\oV =
\left (
\begin{array}{cc} 
\oT^T & \Zero  \\ 
\Zero & \oI  
\end{array} 
\right )
\end{equation}
If both matrix are multiplied in the following way, we notice a very useful
property:
\begin{equation} \label{pgi:UV}
\oU\oV =
\left (
\begin{array}{cc} 
\Zero & \oT  \\ 
\oT^T & \Zero  
\end{array} 
\right )
\end{equation}  
In equation \ref{pgi:UV}, we notice that extra diagonal terms can be expressed
as a product of two matrix. Thus, rewriting the inverted term in equation
\ref{pgi:Kdev}, we have to calculate the following:
\begin{equation} \label{pgi:PUV}
(\oP+\oU\oV)^{-1} 
\end{equation}
This is from that expression that we realise inversion through Woodbury formula
\cite{Zhang05} written below:
\begin{equation}\label{pgi:Wood}
(\oP-\oQ\oR)^{-1} = \oP^{-1} + \oP^{-1}\oQ(\oI-\oR\oP^{-1}\oQ)^{-1}\oR\oP^{-1}
\end{equation}
Changing sign on the left side, we obtain:
\begin{equation} \label{pgi:Woodplus}
(\oP+\oQ\oR)^{-1} = \oP^{-1} - \oP^{-1}\oQ(\oI+\oR\oP^{-1}\oQ)^{-1}\oR\oP^{-1}
\end{equation}
From equation \ref{pgi:Woodplus}, we calculate the inversion needed in formula
\ref{pgi:PUV}.
First, we look after term  $\oP$ in equation \ref{pgi:PUV}. The $\oP$ matrix is
block-diagonal, thus we got :
\begin{equation} 
\oP^{-1} =
\left (
\begin{array}{cc} 
\oP_1^{-1} & \Zero  \\ 
\Zero & \oP_2^{-1}
\end{array} 
\right )
\end{equation}  
Terms of the matrix are supposed to be known and stored \textit{a priori}. Thus
respect to formula \ref{pgi:Woodplus}, we got :
\begin{equation} 
(\oP+\oU\oV)^{-1} = \oP^{-1} - \oP^{-1}\oU(\oI+\oV\oP^{-1}\oU)^{-1}\oV\oP^{-1}
\end{equation}  
We take the following notation:
\begin{equation} \label{pgi:matC}
\oC = \oP^{-1}\oU(\oI+\oV\oP^{-1}\oU)^{-1}\oV\oP^{-1}
\end{equation}  
Thus a new formulation of that gain matrix $\oK$ is obtained:
\begin{equation} \label{pgi:Kperturb}
\oK = \oB\oH(\oP^{-1} - \oC)
\end{equation}  
Within this equation, the term $\oB\oH\oP^{-1}$ corresponds to join together the
column of both matrix  $\oK_1$ and $\oK_2$ from equation \ref{pgi:lossbasis},
that are respectively of size $n \times d_1$ and $n \times d_2$. We will denote
$\oK_0$ this term $\oB\oH\oP^{-1}$, and thus we got:
\begin{equation} 
\oK_0 =
\left (
\begin{array}{cc} 
\oK_1  & \oK_2
\end{array} 
\right )
= \oB\oH\oP^{-1}
\end{equation} 
In the above equation, the $0$ index notice that the gain matrix is build on the
basis of the matrix of each instruments set independently. Thus we can rewrite
equation \ref{pgi:Kperturb} in the following way :
\begin{equation} \label{pgi:Kperturb0}
\oK = \oK_0 - \oB\oH\oC
\end{equation}  
The term $\oB\oH\oC$ in equation  \ref{pgi:Kperturb0} can be then interpreted as
a correction to the matrix $\oK_0$, that is the simple coupling of the two
original $\oK$ matrix. Thus, we have build here a inversion method, where we can
dissociate the contributions of each instrument and the one coming from their
joint use.

Looking more in detail to the matrix $\oC$, defined by equation \ref{pgi:matC},
some points can be noticed. In the term $(\oI-\oV\oP^{-1}\oU)^{-1}$ to be
inverted, we recall that $\oP^{-1}$ matrix is of size $(d_1 + d_2) \times (d_1 +
d_2)$. The $\oV$ and $\oU$ matrices are respectively of size $2 d_2 \times (d_1
+ d_2)$ and $(d_1 + d_2) \times 2 d_2$. The matrix that need do be inverted in
equation \ref{pgi:matC} is then of size $2 d_2 \times 2 d_2$. If the size of the
observation space associated to instrument $2$ is smaller that the size of the
observation space of instrument $1$, it is far better to use this method. Of
course the instrument denoted $1$ and $2$ can be ordered in such a way that $d_1
> d_2$, which ensure that the method always gives benefits.

In the case of the permutation of instrument, we only know the analysis of the
intermediate matrix for the all the instrument but one. Then the calculation of
the analysis for only one instrument need to be performed before using this
method. This correspond on overall to the inversion of an extra $d_2$ size
matrix, which is rather cheap in calculation time. 

\end{document}